\documentclass[aps,draft,showpacs]{revtex4}
\begin{document}

\def\ket#1{|#1\rangle}
\def\bra#1{\langle#1|}
\def\scal#1#2{\langle#1|#2\rangle}
\def\matr#1#2#3{\langle#1|#2|#3\rangle}
\def\keti#1{|#1)}
\def\brai#1{(#1|}
\def\scali#1#2{(#1|#2)}
\def\matri#1#2#3{(#1|#2|#3)}
\def\bino#1#2{\left(\begin{array}{c}#1\\#2\end{array}\right)}

\title{Dynamical and invariant supersymmetry in the fermion
       pairing problem}
\author{Pavel Cejnar$^{1,2,}$}\email{pavel.cejnar@mff.cuni.cz}
\author{Hendrik B. Geyer$^{2,}$}\email{hbg@sun.ac.za}
\affiliation{
$^1$Institute of Particle and Nuclear Physics, Charles University,
V Hole\v sovi\v ck\'ach 2, 180\,00 Prague, Czech Republic\\
$^2$Institute of Theoretical Physics, University of Stellenbosch,
7602 Matieland, South Africa
}
\date{\today}


\begin{abstract}
We argue that fermion-boson mapping techniques represent a natural tool 
for studying many-body supersymmetry in fermionic systems with pairing.
In particular, using the generalized Dyson mapping of a many-level
fermion superalgebra with the SU(2) type of pairing we investigate
two kinds of supersymmetry connecting excitations in the systems
with even and odd particle numbers: dynamical supersymmetry, which 
ensures a unified classification of states for both even and odd 
populations, and invariant supersymmetry with actual degeneracies 
of states within the same supermultiplet. Dynamical supersymmetries 
based on the dynamical algebra of the type U($K$/$2\Omega$) (where 
$K$ and $2\Omega$ denote the number of fermion-pair and 
single-fermion states, respectively) naturally arise in the 
bosonized description of the system. Conditions for invariant 
supersymmetry are determined in a restricted case of bilinear 
supercharge operators.
\end{abstract}

\maketitle

\section{Introduction}
\label{introduction}

Supersymmetry is commonly known as a hypothetical algebraic
scheme in quantum field theory that unifies
internal and space-time symmetries and predicts elementary
particles of matter occurring in boson-fermion doublets
\cite{Wein}.
As shown by Witten \cite{Witten}, the same scheme can be applied
in nonrelativistic quantum mechanics, yielding
a generalized dynamical algebra (superalgebra) which leads to an
analogous doublet structure (except for the unique ground state) in
the spectra of relevant quantum systems.
In the simplest case, the SUSY Hamiltonian $H=a^{\dagger}a+
b^{\dagger}b$ is just the anticommutator of supercharges
$Q=ba^{\dagger}$ and $Q^{\dagger}=ab^{\dagger}$ that change bosonic
excitations ($b^{\dagger}$) into fermionic ones ($a^{\dagger}$) and
vice versa, so that the excited states $\ket{n_a,n_b}$ (where $n_a$
and $n_b$, respectively, are numbers of fermionic and bosonic
quanta) exhibit the characteristic two-fold degeneracy $H\ket{0,n_b}
=H\ket{1,n_b-1}$, coupling boson-like and fermion-like excitations.

Applications of SUSY quantum mechanics soon followed. Analytically
solvable and isospectral sets of potentials were constructed on the
basis of SUSY (see Refs.~\cite{Cooper,Cooper2} and references therein) and
some of these potentials were found relevant for experimental
spectroscopic data of certain atoms and ions. Thus an
(approximate) manifestation of phenomenological supersymmetry was
established in atomic physics \cite{Kostelecky1,Kostelecky2}.
Methods based on the SUSY formalism were also developed in random
matrix theory and applied to systems that exhibit signatures of
quantum chaos \cite{RMT}.

In nuclear physics, the concept of supersymmetry found a natural
application (see e.g.
Refs.~\cite{Iachello2,Balantekin,Balantekin2,Isacker}) in
the framework of the Interacting Boson-Fermion Model
\cite{Iachello3}.
The embedding of the IBFM dynamical algebra of the type
${\rm U}_{\rm B}(K)\otimes{\rm U}_{\rm F}(2\Omega)$ (formed by
$K^2$ bosonic generators $b_i^{\dagger}b_j$ and by
$(2\Omega)^2$ fermionic generators $a^{\dagger}_k a_l$,
where $i,j=1,\dots K$ and $k,l=1,\dots 2\Omega$
enumerate single-boson and -fermion states, respectively) into the
dynamical superalgebra U($K$/$2\Omega$) (with the
mixed generators $b_i^{\dagger}a_l$ and $a^{\dagger}_kb_j$
added) makes it possible to simultaneously describe
low-lying spectra of doublets \cite{Iachello2,Balantekin2} or
quartets \cite{Isacker} of neighboring even and odd nuclei
using a single energy formula with SUSY-based quantum numbers.
(Note that in this paper direct products are used also for
algebras, although rigorously we should speak about a direct
sum of generators associated with the corresponding product of
groups.)

{\em Dynamical supersymmetries\/} are associated with chains
${\cal D}\supset{\cal A}_1\supset{\cal A}_2\supset\dots$ of
possible algebraic decompositions of the dynamical superalgebra
${\cal D}$ and refer to situations when the Hamiltonian $H$ can
be written solely in terms of the Casimir invariants $C({\cal A}_i)$
of (super)algebras in one particular chain, thus implying complete
integrability with a multitude of conservation laws
$[H,C({\cal A}_i)]=0$.

It should be stressed that nuclear supersymmetry, introduced by
Iachello \cite{Iachello2} already in 1980, was historically the
first application of the SUSY ideas in nonrelativistic physics.
Their most detailed verification in nature up to date---in the
recent experimental work by Metz, Jolie {\it et al.\/}
\cite{Metz}---opened up further questions in the SUSY many-body
physics.
Unfortunately, discussions of dynamical supersymmetry on the
phenomenological nuclear-structure level, and its relation to the
notion of SUSY quantum mechanics, have not always clarified the
distinction between them, nor focused on the microscopic basis
of nuclear SUSY.
It is therefore not totally surprising to
find that somewhat negative opinions such as the following one by 't
Hooft \cite{hooft} have been voiced: \lq\lq At first sight, the fact 
that
supersymmetric patterns were discovered in nuclear physics has little
to do with the question of supersymmetry among elementary particles,
but it may indicate that, as the spectrum of particles is getting more
and more complex, some supersymmetric patterns might easily arise,
even if there is no `fundamental' reason for their existence.\rq\rq

The dynamical supersymmetry described theoretically in
Ref.~\cite{Isacker}, and found in experimental data \cite{Metz},
is based on an immediate decomposition
\begin{equation}
{\rm U}(K/2\Omega)\supset
{\rm U}_{\rm B}(K)\otimes{\rm U}_{\rm F}(2\Omega)\supset
\dots
\label{chain}
\end{equation}
of the dynamical superalgebra into a product of the corresponding
bosonic and fermionic algebras, namely on the decomposition
${\rm U}_{\nu}(6/12)\otimes{\rm U}_{\pi}(6/4)\supset
{\rm U}_{{\rm B}\nu}(6)\otimes{\rm U}_{{\rm B}\pi}(6)
\otimes{\rm U}_{{\rm F}\nu}(12)\otimes{\rm U}_{{\rm F}\pi}(4)
\supset\dots$ in the observed case, where $\nu$ and $\pi$,
respectively, stand for neutron and proton realizations of the
above (super)algebras.
This implies that the achieved SUSY description does not really go
beyond an application of the same IBFM Hamiltonian (with fixed
interaction strengths) to a quartet of neighbouring even-even,
even-odd, odd-even, and odd-odd nuclei.
Indeed, since the dynamical-superalgebra irreps with more than one 
fermion in a given nucleus are highly excited, and thus do not mix 
with the low-energy spectrum, the SUSY Casimir invariants can be 
skipped as they only yield the same additive constants to level 
energies in all four nuclei considered \cite{Isacker,Metz}. 
The remaining Casimir invariants (of the embedded algebras) are
precisely those that appear in the standard IBFM description.

As recently demonstrated in Ref.~\cite{Cejnar1}, any collective
superalgebra of fermion-pair and single-fermion operators, linked
to a {\em microscopic\/} perspective of the IBFM, can be naturally
embedded into the phenomenological-type boson-fermionic dynamical
superalgebra U($K/2\Omega$). Here $K$ and $2\Omega$ now denote
the number of pair and single-fermion operators, respectively,
contained in the fermionic collective superalgebra.
Moreover, the same analysis leads to the conclusion that all
microscopically relevant dynamical supersymmetries must
{\em separately\/} conserve the numbers of bosons and fermions, so
that they are always of the form (\ref{chain}).
Dynamical supersymmetry based on the IBFM is therefore always
restricted to the structure (\ref{chain}), and should then probably
only be referred to as dynamical boson-fermion symmetry.

An interesting, though so far mostly hypothetical possibility was,
nevertheless, considered by Jolos and von Brentano \cite{Jolos}.
It is based on the requirement that supercharges $Q_i$, generating
the odd sector of a boson-fermion superalgebra ${\cal I}\subset
{\rm U}(K/2\Omega)$, together with generators of the even sector,
commute with the Hamiltonian, and thus ${\cal I}$ forms
a superalgebra of {\em invariant supersymmetry\/} of the system.
This should be satisfied regardless of whether or not the system
possesses any dynamical supersymmetry of the form (\ref{chain}).
Clearly, the invariant supersymmetry represents principally the
same kind of symmetry as the one introduced by the SUSY quantum
mechanics, although the form of the Hamiltonian can be more general
than simply the anticommutator of supercharges discussed above.
It should be stressed that the bosons considered in this work
are introduced on a purely phenomenological level.

The distinction between the above two SUSY schemes is similar to
the difference between invariant and dynamical symmetries in
standard quantum mechanics: in the former case, all generators
of the given symmetry algebra commute with the Hamiltonian, while
in the latter this is satisfied only for Casimir operators of
a certain chain of subalgebras (the last algebra in the chain
thus represents an invariant supersymmetry).
Dynamical supersymmetry, as presented in nuclear data, implies that
all states in the neighbouring even and odd systems are labelled by
the same set of SUSY quantum numbers, but these states are not
necessarily degenerate.
The invariant supersymmetry, on the other hand, results in actual
degeneracy (at least on a relative scale) of a certain subset of
states, which has not been observed yet.
Needless to say, that in field theory, as well, the SUSY scenario
naturally assumes breaking of the supersymmetry between elementary
bosons and fermions, as large mass differences for particles
within the same supermultiplet must be accommodated.

In this work, we extend the approach of Ref.\cite{Cejnar1} to
study microscopic conditions for supersymmetric schemes
in fermionic many-body systems.
Our present discussion aims to clarify that while dynamical
boson-fermion symmetry seems to be firmly established, and in the final
instance based on the nuclear interaction in situations where some
collective pairs are favoured (together with the utility of
boson-fermion mappings, see Sec.~\ref{mapping}), the direct analogue 
in many-body systems of SUSY quantum mechanics (invariant 
supersymmetry), still requires clarification and may indeed be 
difficult to realise.
This is explicitly demonstrated in Sec.~\ref{supersymmetry}.
Again the precise nature of the interaction on the fermionic level
will of course turn out to be crucial for the appearance of invariant
SUSY.

\section{Bosonized forms of the pairing system}
\label{mapping}

\subsection{Pairing Hamiltonian}
\label{ham}

In this section we present a bosonized form of a many-level
fermionic system with the SU(2) type of pairing, a standard
interaction to capture the essence of superconductivity in solids and
nuclei \cite{BCS}.
We will deal with a set of single-particle states, enumerated by
$k=1,\dots\Omega$, and the respective time-reversal conjugate
states, denoted by ${\bar k}={\bar 1},\dots{\bar\Omega}$, occupied
by an arbitrary number $N_{\rm F}\in[0,2\Omega]$ of fermions.
The fermion creation and annihilation operators corresponding to
individual states are $a_k^{\dagger}$, $a_{\bar k}^{\dagger}$ and
$a_k$, $a_{\bar k}$. The Hamiltonian
\begin{eqnarray}
H_{\rm F} & = & \sum_k E_k\left(a_k^{\dagger}a_k+
a_{\bar k}^{\dagger}a_{\bar k}\right)-\sum_{k,l}
G_{kl}a_{\bar k}^{\dagger}a_k^{\dagger}
a_la_{\bar l}
\label{Hfer}\\
& = & \sum_k E_k N_k-\sum_{k,l}G_{kl}B^{\dagger}_kB_l
\nonumber
\end{eqnarray}
contains as free parameters single-particle energies $E_k$
(states $k$ and ${\bar k}$ are degenerate due to the time-reversal
invariance) and strengths of the pairing interaction $G_{kl}=
G^*_{lk}$.
We can anticipate that the pairing interaction acts only between
fermions in a certain subset ${\cal P}\equiv\{k_1,k_2,\dots k_{\Xi}\}$
(with all $k_i$'s mutually different, $\Xi\leq\Omega$) of states---most
likely states in some interval around the Fermi energy---so that
$G_{kl}\neq 0$ only for $k,l\in{\cal P}$.

The operators
\begin{equation}
B_k^{\dagger}=a_{\bar k}^{\dagger}a_k^{\dagger}
\label{bifer1}
\end{equation}
and $B_k\equiv(B_k^{\dagger})^+=a_ka_{\bar k}$ create and annihilate,
respectively, a pair of fermions ({\em bifermion}) on the shell $k$
(thus there are $\Omega$ such pairs), and together with the
$k$-shell fermion number operator $N_k=a_k^{\dagger}a_k+
a_{\bar k}^{\dagger}a_{\bar k}$ form the SU(2) algebra
associated with each shell; $[B_k^{\dagger},B_k]=N_k-1$.
Thus the Hamiltonian (\ref{Hfer}) leads to the dynamical
algebra $\bigotimes_k{\rm SU(2)}_k$.

For special choices of $E_k$ and $G_{kl}$, the Hamiltonian can be 
rewritten via a smaller number ($K<\Omega$) of pairs,
\begin{equation}
C_{\mu}^{\dagger}=\sum_{k}\chi_{\mu k}a^{\dagger}_{\bar k}a^{\dagger}_k
\label{bifer2}
\end{equation}
and $C_{\mu}\equiv(C_{\mu}^{\dagger})^+$ ($\mu=1,\dots,K$), with the 
coefficients $\chi_{\mu k}$ satisfying the normalization constraint 
$\sum_k\chi_{\mu k}\chi_{\nu k}^*=\delta_{\mu\nu}$.
The condition for the Hamiltonian reads as $H_{\rm F}=H_{\rm F}^0$,
where
\begin{eqnarray}
H^0_{\rm F} & = & {\rm const}+\sum_{\mu,\nu}U_{\mu\nu}
[C^{\dagger}_{\mu},C_{\nu}]
+\sum_{\mu,\nu}V_{\mu\nu}C_{\mu}^{\dagger}C_{\nu}\nonumber\\
& = & {\rm const'}+\sum_k\biggl(\sum_{\mu,\nu}U_{\mu\nu}\chi_{\mu k}
\chi^*_{\nu k}\biggr)N_k+\sum_{k,l}\biggl(\sum_{\mu,\nu}V_{\mu\nu}
\chi_{\mu k}\chi^*_{\nu l}\biggr)B_k^{\dagger}B_l
\label{Hfer2}
\end{eqnarray}
($U_{\mu\nu}=U_{\nu\mu}^*, V_{\mu\nu}=V_{\nu\mu}^*$ are arbitrary 
interaction constants).
Since creation and annihilation operators of {\em any\/} set of
fermion pairs, together with their commutators, closes under
commutation, the dynamical algebra generating the Hamiltonian in
Eq.~(\ref{Hfer2}) can be identified with the algebra associated
with the pairs (\ref{bifer2}) instead of those in Eq.~(\ref{bifer1}).
We call the $C$-bifermions in Eq.~(\ref{bifer2}) {\em collective}.
Note that they couple fermions into pairs in entangled
single-particle states, in contrast to the noncollective
$B$-bifermions (\ref{bifer1}) whose wave functions are separable
(except, of course, unavoidable entanglement due to the
antisymmetrization).
Note that Eq.~(\ref{bifer2}) represents the most general
ansatz for the collective pairs that conserves the form of the
Hamiltonian (\ref{Hfer}).

The possibility to rewrite the pairing Hamiltonian (\ref{Hfer}) via
a collective algebra with a small number of $C$-bifermions
is not very exceptional.
In general, it requires to satisfy a set of equations for unknown
coefficients $\chi_{\mu k}$ and for interaction parameters 
$U_{\mu\nu}$, $V_{\mu\nu}$, cf. Eqs.~(\ref{Hfer}) and (\ref{Hfer2}).
Dimensionality considerations suggest that some solutions should
typically exist if $K/\Omega\geq\sqrt{1+1/\Omega+(\Xi/\Omega)^2}-1$,
which for $\Xi\ll\Omega$ indeed leads to $K\ll\Omega$.
Let us stress here the importance of the requirement that also
single-particle terms in $H_{\rm F}$, not only the interactions,
are expressed via the collective algebra, see Eq.~(\ref{Hfer2}).
For instance, the constancy of interaction strengths in
Eq.~(\ref{Hfer}), $G_{kl}=G$, would not imply that the simple
SU(2) collective algebra generated by the single bifermion
$C^{\dagger}=\sum_{k\in{\cal P}}a^{\dagger}_ka^{\dagger}_{\bar k}/
\sqrt{\Xi}$, is the dynamical algebra of the problem, unless all
the single-particle states subject to interactions are degenerate,
i.e., $E_k=E$ for $k\in{\cal P}$:
\begin{equation}
H_{\rm F}^0=\sum_{k{\not\in}{\cal P}} E_kN_k+
EN_{\cal P}-G\,\Xi\,C^{\dagger}C\ ,
\label{HC}
\end{equation}
where $N_{\cal P}=\sum_{k\in{\cal P}}N_k$.

Another example of the use of a collective algebra concerns a system
of fermions on $K$ single-$j$ shells, $j=j_1, j_2, \dots, j_K$, with
a Hamiltonian expressed just via interactions of $J=0$ pairs on
individual shells, $S_{\mu}^{\dagger}=(a^{\dagger}_{j_{\mu}}
a^{\dagger}_{j_{\mu}})^{(0)}$:
\begin{equation}
H^0_{\rm F}=\sum_{\mu}E_{\mu}N_{\mu}-\sum_{\mu,\nu} 
V_{\mu\nu} S_{\mu}^{\dagger}S_{\nu}
\ ,\label{HS}
\end{equation}
($E_{\mu}$ are the $j$-shell energies, $N_{\mu}$ the corresponding
occupation-number operators, and $V_{\mu\nu}$ the interaction
strengths).
Again, the Hamiltonian (\ref{HS}) is of the general form
(\ref{Hfer}), with the conjugate states $k$ and ${\bar k}$
corresponding to opposite projections $+m_{\mu}$ and $-m_{\mu}$ of
the angular momentum for the same level $\mu$.

\subsection{Mapping via noncollective pairs}
\label{mapp1}

The dynamical algebra of the fermion pairing problem can be translated
into boson language with available boson mapping
techniques \cite{Klein}, mostly explored for even fermion systems.
If both even and odd particle numbers are to be described simultaneously,
the algebra of fermion pairs must be extended to take into account also
the odd degrees of freedom.
This can most naturally be done by adding the single-fermion creation
and annihilation operators to the given set of bifermions.
The resulting collection of operators then forms a superalgebra
with the algebra of pair operators as a subalgebra.

As shown in Refs.~\cite{Dobaczewski,Navratil1,Navratil2},
partial bosonization of this extended superalgebra can be
achieved by a generalized Dyson mapping which, in the case of the
superalgebra based on the above noncollective pairs (\ref{bifer1}),
leads to
\begin{eqnarray}
B_k^{\dagger} & \mapsto & b_k^{\dagger}(1-n_k-{\cal N}_k)\ ,
\label{map1}\\
B_k           & \mapsto & b_k\ ,
\label{map2}\\
N_k           & \mapsto & 2n_k+{\cal N}_k
\label{map3}
\end{eqnarray}
for the even sector, and
\begin{eqnarray}
a_k^{\dagger} & \mapsto & b_k^{\dagger}\alpha_{\bar k}+\alpha_k^{\dagger}
\frac{1-{\cal N}_k-n_k}{1-{\cal N}_k}-\alpha_k^{\dagger}
\alpha_{\bar k}^{\dagger}\alpha_{\bar k}\frac{1-{\cal N}_k-n_k}
{(2-{\cal N}_k)(1-{\cal N}_k)}
\label{map4}\\
a_k            & \mapsto & \alpha_k+\alpha_{\bar k}^{\dagger}b_k
\frac{1}{1-{\cal N}_k}+\alpha_k^{\dagger}\alpha_{\bar k}^{\dagger}
\alpha_kb_k\frac{1}{(2-{\cal N}_k)(1-{\cal N}_k)}
\label{map5}
\end{eqnarray}
for the odd sector \cite{Navratil2,Cejnar1}.
In the right-hand side images, $b_k^{\dagger}$
($b_k$) creates (annihilates) a {\em boson\/} of the $k$th type
and $n_k=b_k^{\dagger}b_k$, while
$\alpha_k^{\dagger}$ (or $\alpha_k$) create (or annihilate)
{\em ideal fermions\/} in the states $k$ (similarly for ${\bar k}$),
and ${\cal N}_k=\alpha_k^{\dagger}\alpha_k+\alpha_{\bar k}^{\dagger}
\alpha_{\bar k}$. These bosons and ideal fermions are kinematically
independent, i.e.\ all boson operators commute with all fermion
operators.
The images of $a^{\dagger}_{\bar k}$ and $a_{\bar k}$ can be deduced
from Eqs.~(\ref{map4}) and (\ref{map5}), respectively, after the
$k\leftrightarrow{\bar k}$ exchange of indices in the fermionic
operators and inverting signs of the first term in Eq.~(\ref{map4})
and of the second and third term in Eq.~(\ref{map5}).

The term \lq ideal fermion\rq\ is used to distinguish the fermion-type
particles
resulting from the mapping (where they appear as  necessary
ingredients
of any superalgebraic extension of boson mapping techniques) from the
\lq real\rq\ fermions in the original formulation of the problem. The
Bogoliubov-Valatin like structure of Eqs.~(\ref{map4}) and
(\ref{map5}) suggests a physical interpretation
of ideal fermions as generalized quasiparticles
\cite{Suzuki}.
Anticipating discussions that follow below, we already point out
here that possible divergences associated with the denominators in
these formulas simply do not appear in the physical subspace.
Furthermore
the loss of symmetry between creation and annihilation operators under
Hermitian conjugation (a typical feature of the Dyson type of
mappings) is accounted for in the calculation of observables within
the general Dyson framework, see Refs.\cite{Klein,Cejnar1}.

The mapped Hamiltonian reads as follows:
\begin{equation}
H_{\rm B}=\sum_k E_k(2n_k+{\cal N}_k)-\sum_{k,l}
G_{kl}b_k^{\dagger}b_l(1+\delta_{kl}-n_k-{\cal N}_k)\ .
\label{Hbos}
\end{equation}
It is clear that this Hamiltonian conserves both the total number of
bosons, $N_{\rm B}=\sum_k n_k$, and of ideal fermions, ${\cal N}_{\rm
F} =\sum_k{\cal N}_k$, so that by considering a fixed total number of
{\em real\/} fermions, $N_{\rm F}=\sum_k N_k$, we also fix the sum
$2N_{\rm B}+{\cal N}_{\rm F}$ to the given value $N_{\rm F}$; see
Eq.~(\ref{map3}).

The Dyson mapping of states can be trivially deduced from
Eqs.~(\ref{map1})--(\ref{map5}). Four real-fermion
basis states in each $k$-subspace yield the ideal boson-fermion
images as follows: $\ket{0}\mapsto\keti{0}$, $a_k^{\dagger}\ket{0}
\mapsto\alpha_k^{\dagger}\keti{0}$, $a_{\bar k}^{\dagger}\ket{0}
\mapsto\alpha_{\bar k}^{\dagger}\keti{0}$, $B_k^{\dagger}\ket{0}
\mapsto b_k^{\dagger}\keti{0}$; we denote real- and ideal-space
vectors by angular and circled bras/kets, respectively, as in case of
real and ideal vacua $\ket{0}$ and $\keti{0}$. All the states
in the ideal Hilbert space that are {\em not\/} linear combinations of
the images just given, are spurious. In particular, the components
containing in any $k$-shell more than one ideal particle
(whether bosons or fermions, or both) are nonphysical. We see,
therefore, that the present model of pairing allows
us to write the projector to the physical subspace in the following
{\em explicit\/} form:
\begin{equation}
P_{\rm ph}=\prod_k\left(P_{n_k=0}P_{{\cal N}_k=0}+P_{n_k=0}
P_{{\cal N}_k=1}+P_{n_k=1}P_{{\cal N}_k=0}\right)\ ,
\label{proj}
\end{equation}
where $P$'s on the right-hand side represent projectors onto the ideal
subspaces with the given number of ideal particles (bosons or fermions)
of the $k$th type.

Because of the nonunitary character of the Dyson mapping, the
boson-fermionic Hamiltonian in Eq.~(\ref{Hbos}) (as well as images of 
other physical observables) is non-Hermitian with respect to the 
standard boson-fermion Fock space inner product. 
However, it is well known that Dyson mappings lead to
so-called quasi-Hermitian structures which are consistent with
standard quantum mechanics, and in particular guarantees real
eigenvalues for observables (see Refs.\cite{Scholtz,Cejnar1,Kim}).
Although explicit hermitization has been achieved
in some particular cases \cite{Kim}, it seems that this procedure
will generally introduce higher order
many-body interaction terms into the Hamiltonian.
Fortunately, in the present case, it can be easily checked
that the Hamiltonian in Eq.~(\ref{Hbos}) {\em is\/} already Hermitian
{\em within\/} the physical subspace, i.e., it satisfies
$H_{\rm B}P_{\rm ph}=H'_{\rm B}P_{\rm ph}$, where
\begin{eqnarray}
H'_{\rm B} & = & \sum_k E_k(2n_k+{\cal N}_k)-
\sum_{k,l}G_{kl}b_k^{\dagger}b_l
\nonumber\\
& = & \sum_k E_k{\cal N}_k+\sum_k(2E_k-G_{kk})
n_k-\sum_{k\neq l}G_{kl}
b_k^{\dagger}b_l
\label{Hbos2}
\end{eqnarray}
is manifestly Hermitian. We see that the Hamiltonian (\ref{Hbos2})
contains no interaction terms---it is just a combination of bosonic
and fermionic mean fields---although both boson-boson and boson-fermion
interactions were present in the original mapped Hamiltonian
(\ref{Hbos}). This implies that the interactions in Eq.~(\ref{Hbos})
do not affect physical states.

The Hamiltonian (\ref{Hbos2}) has the same form as the one in
Eq.~(3) of Ref.~\cite{Celeghini}. Note that while in their treatment
of pairing the authors of the cited work introduce boson-like
particles, so-called cooperons, by modifying fundamental
anticommutation relations of real fermions, our bosons result from
the mapping of a conventional multifermionic superalgebra. In
contrast to cooperons, the bosons in our case obey ordinary
commutation relations, but for the limitations concerning the
physical subspace they are in fact {\em hard-core bosons}, i.e.,
bosons with occupation numbers restricted to 0 and 1.

In spite of the single-particle form of the Hamiltonian
(\ref{Hbos2}), its diagonalization in terms of some new bosons
$d_l^{\dagger}=\sum_k\beta_{lk}b_k^{\dagger}$ and
identification of the ground state for even systems with
a condensate
\begin{equation}
\keti{{\rm cond}}\propto\biggl(\sum_k\eta_kb_k^{\dagger}\biggr)
^{N_{\rm B}}\keti{0}
\label{cond}
\end{equation}
(where $N_{\rm B}=N_{\rm F}/2$ and $\eta_k=\beta_{l_0k}$, with
$l_0$ denoting the $d$-boson with minimum energy),
are not physically allowed operations. Indeed, such a procedure
does not keep under control physicality of the transformed states;
the condensate state has a finite spurious admixture as it
contains terms with more than one boson on a given $k$-shell
(cf. Ref.~\cite{Dobes}).
The overlap of the state (\ref{cond}) with the physical subspace
of the ideal space is given by the following expression:
\begin{equation}
\frac
{|\matri{{\rm cond}}{P_{\rm ph}}{{\rm cond}}|^2}
{\scali{{\rm cond}}{{\rm cond}}}
=
\frac
{|\sum_{k_1\neq k_2\neq\dots\neq k_{N_{\rm B}}}
\eta_{k_1}\eta_{k_2}\dots\eta_{k_{N_{\rm B}}}|^2}
{\left(\sum_k|\eta_k|^2\right)^{N_{\rm B}}N_{\rm B}!}
\ ,
\label{overlap}
\end{equation}
which is less than unity for $1<N_{\rm B}\leq N_{\rm F}/2$ and
any nontrivial set of $\eta$'s. At the same time,
$P_{\rm ph}\keti{\rm cond}$ is generally {\em not\/} an eigenstate
of $H'_{\rm B}$, nor of $H_{\rm B}$.

On the other hand, properties of the ground state of an even system
can be estimated from the bosonic equivalent of the BCS approximation
\cite{BCS}, using  the image of the BCS state $\ket{{\rm BCS}}=
\prod_k(u_k+v_kB_k^{\dagger})\ket{0}$, namely
\begin{equation}
\keti{{\rm BCS}}=\prod_k(u_k+v_kb_k^{\dagger})\keti{0}=P_{\rm ph}
\exp\sum_k\left(\ln u_k+\frac{v_k}{u_k}b_k^{\dagger}\right)\keti{0}
\label{coherent}
\end{equation}
with the normalization condition $u_k^2+v_k^2=1$ for each $k$.
The trial wave function (\ref{coherent}) yields the following
energy functional:
\begin{equation}
\matri{{\rm BCS}}{H'_{\rm B}-2\lambda N_{\rm B}}{{\rm BCS}}=
\sum_k(2E_k-G_{kk}-2\lambda)v_k^2
-\sum_{k\neq l}G_{kl}u_kv_ku_lv_l\ ,
\label{Efunc}
\end{equation}
(with $\lambda$ denoting a Lagrange multiplier), the same expression as
in the standard BCS approximation.

We see that the bosonic BCS state (\ref{coherent}) is just an ordinary
Glauber coherent state projected onto the physical subspace (while the
original BCS state is a generalized coherent state of the fermion
dynamical group \cite{Zhang}).
If the exponential on the right-hand side in Eq.~(\ref{coherent}) is
projected onto a fixed number of particles (instead of the projection
onto the physical space) one would get exactly the condensate state
(\ref{cond}) with $\eta_k=v_k/u_k$, but---as discussed above---this
would not be a physically justified procedure.
Instead, the use of $P_{\rm ph}$ leads to the cutoff of spurious
components, which also modifies the normalization factor in front of
$\exp\sum_k\eta_kb_k^{\dagger}\keti{0}$ in Eq.~(\ref{coherent}) with
respect to the standard bosonic coherent state.

\subsection{Mapping via collective pairs}
\label{mapp2}

If the pairing Hamiltonian can be expressed in the form
(\ref{Hfer2}), one can perform the fermion-boson
mapping via the corresponding collective pairs.
This is particularly simple in both the SU(2) special cases
discussed above, see Eqs.~(\ref{HC}) and (\ref{HS}):
The $C$-bifermion from Eq.~(\ref{HC}) yields a $c$-boson,
$C^{\dagger}\mapsto c^{\dagger}[1-(n+{\cal N})/\Xi]$, while
the $S$-bifermions from Eq.~(\ref{HS}) map onto $s$-bosons
as follows: $S^{\dagger}_{\mu}\mapsto s^{\dagger}_{\mu}
\left[1-(n_{\mu}+{\cal N}_{\mu})/(j_{\mu}+\frac{1}{2})\right]$.
Note that the bifermion annihilation operators are in both cases
mapped trivially just onto the corresponding boson annihilation
operators, and the fermion number operators from the collective
algebras onto analogous combinations as in Eq.~(\ref{map3}).
For instance, $N_{\mu}\mapsto 2n_{\mu}+{\cal N}_{\mu}$, where 
$n_{\mu}$ and ${\cal N}_{\mu}$ stand for the boson and 
ideal-fermion number operators, respectively, of the $\mu$th 
level in Eq.~(\ref{HS}).
It is clear that in this case the physical space is not restricted
to the $s$-boson occupation numbers $n_{\mu}=0$ and 1, but contains
states with $n_{\mu}$ up to $j_{\mu}+\frac{1}{2}$.
The single-fermion operators, that in both of the discussed cases
supplement the bifermion algebra to yield the corresponding
collective superalgebra, are also mapped analogously as in the
previous section, cf. expressions in Refs.~\cite{Navratil2,Cejnar1}

The collective algebra of the bifermions (\ref{bifer2}) is not
generally of the SU(2) type because different collective pairs,
$C_{\mu}^{\dagger}$ and $C_{\nu}^{\dagger}$, $\mu\neq\nu$, may 
contain components with the same $k$.
Nevertheless, using the formalism described in
Refs.~\cite{Navratil1} and \cite{Cejnar1}, the Dyson bosonic images
can be easily constructed:
\begin{eqnarray}
C_{\mu}^{\dagger} & \mapsto & c_{\mu}^{\dagger}-
\sum_{\nu}c_{\nu}^{\dagger}\biggl(\sum_k\chi^*_{\mu k}\chi_{\nu k}
\,{\cal N}_k+\sum_{\omega,\pi}\sum_k
\chi_{\mu k}\chi^*_{\nu k}\chi^*_{\omega k}\chi_{\pi k}
\,c_{\omega}^{\dagger}c_{\pi}\biggr)\ ,
\label{map11}\\
C_{\mu} & \mapsto & c_{\mu}\ ,
\label{map22}\\
\left[C^{\dagger}_{\mu},C_{\nu}\right] & \mapsto & -\delta_{\mu\nu}+
\sum_k\chi_{\mu k}\chi^*_{\nu k}\,{\cal N}_k+2\sum_{\omega,\pi}\sum_k
\chi_{\mu k}\chi^*_{\nu k}\chi^*_{\omega k}\chi_{\pi k}\,
c_{\omega}^{\dagger}c_{\pi}\ .
\label{map33}
\end{eqnarray}
Now $c_{\mu}^{\dagger}$ and $c_{\mu}$ create and annihilate, respectively,
the boson corresponding to the $\mu$th collective bifermion.
Note, however, that single-particle creation and annihilation
operators generating the odd sector of the general collective
superalgebra are more difficult to determine than in the
SU(2)-based special cases \cite{Cejnar1}.

It should be stressed that although the collective $C$-bifermions
are just linear combinations of the noncollective ones,
$C_{\mu}^{\dagger}=\sum_k\chi_{\mu k}B^{\dagger}_k$, the respective
collective and noncollective bosons are {\em not\/} connected by
the same linear relation.
Note that the two kinds of boson operators can be viewed to act 
formally in two
different Hilbert spaces, so that their comparison requires to
introduce an operator $T$ which transforms the physical space
of the collective mapping onto the physical space of noncollective
mapping.
Eq.~(\ref{map22}) and linearity of all mapping procedures
trivially yield $Tc_{\mu}T^{-1}\equiv{\tilde c}_{\mu}=
\sum_k\chi^*_{\mu k}b_k$;
it means that the linear relation between $C$- and $B$-bifermions
is preserved for the respective boson annihilation operators.
For the creation operators, however, the situation is much more
complicated, as the application of $T$ on the appropriate
linear superposition of the mapping (\ref{map11}) results in
a difficult selfconsistent relation containing combinations of
operators ${\tilde c}_{\mu}^{\dagger}\equiv Tc_{\mu}^{\dagger}T^{-1}$,
$b_k^{\dagger}$, $b_l$, and also ${\tilde{\cal N}_k}\equiv
T{\cal N}_kT^{-1}$.
This relation cannot be generally satisfied by any ansatz of the
type ${\tilde c}^{\dagger}_{\mu}=\sum_k\chi'_{\mu k}b_k^{\dagger}$, 
where $\chi'_{\mu k}$ would represent some unknown coefficients, and 
thus the correspondence between both kinds of mapping is not linear.

\section{Supersymmetry of the pairing Hamiltonian}
\label{supersymmetry}

\subsection{Possible SUSY schemes}
\label{tydyt}

As explicitly demonstrated by the Hamiltonian in Eq.~(\ref{Hbos2}),
the total number of particles is a conserved quantity in the system
of bosons and ideal fermions obtained by the mapping
(\ref{map1})--(\ref{map5}).
The same holds true also for any collective mapping, see
Eqs.~(\ref{map11})--(\ref{map33}) (neither projection to the
physical space, nor the Hermitization can spoil this property
\cite{Cejnar1}).
Therefore, the dynamics of the mapped system does not rule out
the use of the U($K$/$2\Omega$) superalgebra as the spectrum
generating (dynamical) superalgebra.
In fact, the superalgebra of real fermions and bifermions, either
collective or noncollective, is mapped {\em into\/} (but generally
not onto!) this superalgebra.

For the noncollective mapping, for instance, $K=\Omega$ and the
even sector of U($\Omega$/2$\Omega$) is formed by generators
$b_k^{\dagger}b_l$, $\alpha^{\dagger}_k\alpha_l$,
$\alpha^{\dagger}_k\alpha_{\bar l}$, and their Hermitian conjugates
(all together $5\Omega^2$ operators), while the odd sector is
generated by $b_k^{\dagger}\alpha_l$, $b_k^{\dagger}\alpha_{\bar l}$,
and the conjugates (all together $4\Omega^2$) operators.
We see, however, that the odd sector is not used in the construction
of the Hamiltonian, so that ${\rm U}_{\rm B}(\Omega)\otimes
{\rm U}_{\rm F}(2\Omega)$ can equally well be chosen as the
dynamical algebra.
This means that the decomposition in Eq.~(\ref{chain}) is applicable, 
which preselects possible dynamical symmetries of the system.
Realizing that all these symmetries are given by standard
decompositions of the ${\rm U}_{\rm B}(K)\otimes{\rm U}_{\rm F}
(2\Omega)$ algebra into chains of embedded subalgebras, we skip
here their explicit discussion.
In general, conditions for any specific dynamical symmetry, given
by a chain ${\cal D}\supset{\cal A}_1\supset{\cal A}_2\supset
\dots$ of decomposition of the boson-fermion dynamical algebra
${\cal D}$, are obtained from the set of constraints
$[H_{\rm B},C({\cal A}_i)]=0$ required to hold in the physical 
space.
This naturally yields specific (for each symmetry) sets of
equations for parameters $E_k$ and $G_{kl}$ of the Hamiltonians
(\ref{Hbos2}) and (\ref{Hfer}).

Although---as we just saw---the odd sector of the dynamical
superalgebra U$(K/2\Omega)$ can be ruled out, it is also clear
that the SUSY quantum number $\aleph=N_{\rm B}+{\cal N}_{\rm F}$
still classifies all eigenstates of the mapped many-body
Hamiltonian.
Conservation of the total particle number is a common feature of
the generalized Dyson mapping.
Let us stress that this simple conclusion was not at all clear
in the early days of nuclear supersymmetry, especially in view
of the fact that the real fermion number maps onto the number of
ideal fermions plus {\em twice\/} the number of bosons,
$N_{\rm F}\mapsto{\cal N}_{\rm F}+2N_{\rm B}$, which excludes the
particle-number conservation law for real fermions from the
explanation of the conservation of $\aleph$.
Indeed, as shown in Ref.~\cite{Cejnar1}, the conservation law
for $\aleph$ (as well as the separate conservation of $N_{\rm B}$
and ${\cal N}_{\rm F}$) follows from a consistent choice of the
real-fermion dynamical algebra, which must represent the
interaction as well as single-particle terms of the fermion
Hamiltonian.
Therefore, it is only the possibility of fully {\em algebraic
formulation\/} of the fermion pairing problem that automatically
creates the U($K$/$2\Omega$)-based boson-fermion description,
with all potential dynamical-supersymmetry chains included.

Potential {\em invariant\/} supersymmetries of the pairing Hamiltonian 
(\ref{Hfer}) also depend on the the energies $E_k$ and interaction 
strengths $G_{kl}$.
On the fermion level, all such supersymmetries are described in the 
framework of the most general superalgebra of multifermion operators,
\begin{equation}
A^{k_1,\dots k_n,{\bar k}_1,\dots {\bar k}_{\bar n}}
_{k'_1,\dots k'_{n'},{\bar k'}_1,\dots {\bar k'}_{\bar n'}}=
a^{\dagger}_{k_1}\dots a^{\dagger}_{k_n}
a^{\dagger}_{{\bar k}_1}\dots a^{\dagger}_{{\bar k}_{\bar n}}
a_{k'_1}\dots a_{k'_{n'}}a_{{\bar k'}_1}\dots a_{{\bar k'}_{n'}}\ ,
\label{A}
\end{equation}
where a given element belongs to the even or odd sector, respectively, 
according to whether the difference $\Delta N_{\rm F}=(n+{\bar n})-
(n'+{\bar n'})$ is even or odd.
Operators of the even sector transform states with even and odd fermion numbers
separately, while operators of the odd sector interconnect even and odd 
populations.
The invariant SUSY takes place if the Hamiltonian (\ref{Hfer}) commutes with 
any of the operators (\ref{A}) belonging to the odd sector, which implies the
existence of a degenerate supermultiplet of states differing by a given odd
number of $\Delta N_{\rm F}$ fermions.

In the Hilbert space of bosons and ideal fermions created by the mapping 
(\ref{map1})--(\ref{map5}), similarly, one can consider the superalgebra of 
operators
\begin{equation}
B^{k_1,\dots k_n,{\bar k}_1\dots {\bar k}_{\bar n},l_1,\dots l_m}
_{k'_1,\dots k_{n'},{\bar k'}_1,\dots {\bar k'}_{\bar n'},l'_1\dots l'_{m'}}=
{\cal A}^{k_1,\dots k_n,{\bar k}_1,\dots {\bar k}_{\bar n}}
_{k'_1,\dots k_{n'},{\bar k'}_1,\dots {\bar k'}_{\bar n'}}
b^{\dagger}_{l_1}\dots b^{\dagger}_{l_m}b_{l'_1}\dots b_{l'_{m'}}\ ,
\label{B}
\end{equation}
where ${\cal A}^{k_1,\dots}_{k'_1,\dots}$ is defined in the same way as 
$A^{k_1,\dots}_{k'_1,\dots}$ in Eq.~(\ref{A}), but with the real-fermion
operators substituted by the corresponding ideal-fermion operators.
As in the previous case, even and odd sectors of the superalgebra (\ref{B}) 
are distinguished according to the difference $\Delta {\cal N}_{\rm F}$ 
between the numbers of creation and annihilation operators in 
${\cal A}^{k_1,\dots}_{k'_1,\dots}$.

Although the mapping in Sec.~\ref{mapp1} was performed only for a small subset 
of the above general fermionic superalgebra, the images of operators (\ref{A}) 
in the physical space can be constructed as the corresponding products of the
images in Eqs.~(\ref{map4}) and (\ref{map5}) \cite{Dobaczewski}.
The superalgebra (\ref{A}) is thus mapped to series of operators (\ref{B}),
i.e., onto the given superalgebra acting in the space of bosons and ideal
fermions.
Trivially, even and odd sectors of the superalgebra (\ref{A}) are not mixed 
by the mapping---the image of an element with a given value of 
$\Delta N_{\rm F}$ has $\Delta{\cal N}_{\rm F}+2\Delta N_{\rm B}=
\Delta N_{\rm F}$, where $\Delta N_{\rm B}=m-m'$, see Eq.~(\ref{B}).
Taking $\Delta N_{\rm F}=+1$, for example, we get an image with terms
corresponding to $(\Delta{\cal N}_{\rm F},\Delta N_{\rm B})=(+1,0), (-1,+1),
(+3,-1), (-3,+2), \dots$

This expansion would hardly be tractable in its general form, but fortunately,
it can be considerably simplified if states with more than one ideal fermion 
are physically irrelevant.
Indeed, such states correspond to real-fermionic states with broken pairs which
are supposed to have relatively high excitation energies. 
Under this restriction, the invariant-SUSY supercharges connect states
in doublets with $N_{\rm F}$ and $N_{\rm F}+1$ fermions, and can be 
constructed just from the $(\Delta{\cal N}_{\rm F},\Delta N_{\rm B})=(+1,0)$
and $(-1,+1)$ terms.
The first term applies in case that $N_{\rm F}$ corresponding to the left
system in the doublet is even, the second if $N_{\rm F}$ is odd.
To make contact with the standard notion of a supercharge, we here only focus 
on the second term that changes a fermion into a boson (or vice versa). 
It reads as
\begin{equation}
{\cal Q}^{\dagger}=\sum_{k,l}\ b^{\dagger}_k\,
f_{kl}(b_i^{\dagger}b_j,\alpha^{\dagger}_m\alpha_n)
\,\left(q_{kl}\alpha_l+{\bar q}_{kl}\alpha_{\bar l}\right) \ ,
\label{gensup}
\end{equation}
where $f_{kl}$ are arbitrary functions of operators conserving separately 
the numbers of bosons and fermions, and $q_{kl}$ and ${\bar q}_{kl}$ are 
some coefficients.

In spite of the above restrictions, the commutator of the 
supercharge (\ref{gensup}) with the boson-fermion
Hamiltonian still remains completely unknown without a further specification 
of $f_{kl}$.
This makes the general theoretical determination of invariant
supersymmetries in fermionic systems a rather nontrivial task,
where no analytic insight seems available so far.
Nevertheless, the mapping procedure, as outlined in
Sec.~\ref{mapping}, provides a natural framework for the
classification and analysis of various special cases.
In the next paragraph, for example, we determine conditions for the
invariant SUSY generated by supercharges (\ref{gensup}) with
$f_{kl}=1$.

\subsection{Invariant SUSY with bilinear supercharges}
\label{invariant}

We will look for conditions that the mapped Hamiltonian from
Sec.~\ref{mapp1} must satisfy to commute with the simplest bilinear
supercharges of the form
\begin{equation}
Q^{\dagger}=\sum_{k,l}b_k^{\dagger}(q_{kl}\alpha_l+
{\bar q}_{kl}\alpha_{\bar l})
\equiv \sum_{k,l}b_k^{\dagger}{\hat\alpha}_{kl}\ ,
\label{supercharge}
\end{equation}
where $b^{\dagger}_k$ and $\alpha_l$ (or $\alpha_{\bar l}$)
are noncollective bosons and ideal fermions, respectively, from
Eqs.~(\ref{map1}) and (\ref{map5}).
If the supercharge is assumed to commute with the time reversal,
the coefficients $q_{kl}$ and ${\bar q}_{kl}$ should be set
equal (up to a phase factor), but this we do not generally require 
here.

The assumption $[H'_{\rm B},Q^{\dagger}]=0$ ($=[Q,H'_{\rm B}]^+$)
leads to the following equation to be valid for any $k$ and $l$:
\begin{equation}
(2E_k-E_l-G_{kk})q_{kl}
-\sum_{m\neq k}G_{km}q_{ml}=0\ ,
\label{cond1}
\end{equation}
and the same also with the change of $q$'s to ${\bar q}$'s. 
For a fixed value of $l$, the condition for the existence of
nontrivial solutions $q_{kl}$ and ${\bar q}_{kl}$ reads as
\begin{eqnarray}
{\rm det}
\left(
\begin{array}{cccc}
G_{11}-2E_1+E_l & G_{12} & G_{13} & \dots \\
G_{21} & G_{22}-2E_2+E_l & G_{23} & \dots \\
G_{31} & G_{32} & G_{33}-2E_3+E_l & \dots \\
\dots  & \dots & \dots & \dots
\end{array}
\right)
=0.
\label{det}
\end{eqnarray}
This equation has rather clear physical interpretation: it says 
that diagonalization of the bosonic part of the single-particle Hamiltonian 
(\ref{Hbos2}) gives one of the eigenvalues equal to the fermionic 
energy $E_l$. Mutual conversions of the corresponding boson and 
fermion thus do not change the total energy.
One can apply Eq.~(\ref{det}) simultaneously to a certain subset 
of fermionic states, $l\in{\cal S}$, which is equivalent
with taking $q_{kl}={\bar q}_{kl}=0$ for $l\not\in{\cal S}$.
The supersymmetry then concerns only a part of the spectrum.

An important special case of the above condition is obtained if 
one assumes supercharges with diagonal matrices of coefficients,
$q_{kl}=q_k\delta_{kl}$ and ${\bar q}_{kl}={\bar q}_k
\delta_{kl}$. Eq.~(\ref{cond1})  
then leads to the following simple constraints: $G_{kl}=0$ for
$k\neq l\in{\cal S}$ and $E_k=G_{kk}$ for $k\in{\cal S}$, the 
coefficients $q_k$ and ${\bar q}_k$ being arbitrary for
$k\in{\cal S}$ and zero otherwise. We thus have $H'_{\rm B}=H'
_{\rm SUSY}$ with
\begin{equation}
H'_{\rm SUSY}=\sum_{k\in{\cal S}}E_k(n_k+{\cal N}_k)
+\sum_{k\not\in{\cal S}}E_k(2n_k+{\cal N}_k)
-\sum_{k,l\not\in{\cal S}}G_{kl}b_k^{\dagger}b_l\ .
\label{susy}
\end{equation}
The interpretation of this solution is the same as above, with
only the difference that the bosonic part of the Hamiltonian
(\ref{Hbos2}) is already supposed to be diagonalized {\it a priori}, 
allowing for the diagonal constraint on the supercharge coefficients.
By inserting the above constraints into the original Hamiltonian 
(\ref{Hfer}), one can verify the existence of degenerated 
supermultiplets with $N_k=1$ and 2 (for $k\in{\cal S}$) directly 
on the fermionic level.

The Hamiltonian (\ref{susy}) is trivially invariant under the set
of transformations $b^{\dagger}_k\alpha_k$, $b^{\dagger}_k\alpha
_{\bar k}$, $\alpha^{\dagger}_k\alpha_{\bar k}$,
$\alpha^{\dagger}_k\alpha_{\bar k}$, $\alpha^{\dagger}_{\bar k}
\alpha_{\bar k}$, $b^{\dagger}_k b_k$ (and their Hermitian
conjugates) with $k\in{\cal S}$.
These are generators of the ${\cal I}=\bigotimes_{k\in{\cal S}}
{\rm U}_k(1/2)$  superalgebra, which thus forms the invariant-SUSY 
superalgebra of the problem.
This is so in spite of the fact that the dynamical algebra of
$H'_{\rm SUSY}$ in Eq.~(\ref{susy}) is just an ordinary algebra
${\cal D}\subset {\rm U}_{\rm B}(\Omega)\otimes 
{\rm U}_{\rm F}(2\Omega)$ where seemingly no supersymmetry is involved.
In view of the discussion above, the SUSY Hamiltonian obtained from
the general solution of Eq.~(\ref{det}) in case of nondiagonal 
matrices of supercharge coefficients receives the same algebraic 
interpretation, but with the original bosons $b^{\dagger}_k$ 
substituted by the new ones, $d^{\dagger}_k$, resulting from the 
diagonalization of the bosonic Hamiltonian.
Note that the above superalgebraic scheme represents a direct
generalization of the minimal SUSY scheme \cite{Witten}, where the 
invariant-SUSY superalgebra sl(1/1) is formed only by operators 
$Q^{\dagger}$, $Q$, and $H=\{Q^{\dagger},Q\}$ \cite{Cooper,Cooper2}.
In the present case, in particular, the relevant part of the 
Hamiltonian is not just a supercharge anticommutator (or a sum
of such terms), but a linear combination of Casimir invariants 
of $\cal I$.

Let us turn, at last, to the crucial question of physicality
of the above supersymmetric transformations. It was so far completely 
ignored in this section but it is indeed very relevant since the
supercharge operator (\ref{supercharge}) can transform some of
the physical states out of the physical space given by the projector 
in Eq.~(\ref{proj}). In particular, the supercharge operator acting
within the physical space may produce a nonphysical state with 
$(n_k,{\cal N}_k)=(1,1)$ or $(2,0)$, this being not generally 
excluded---in case of nondiagonal matrices of supercharge 
coefficients---by the action of $\alpha_l$ or $\alpha_{\bar l}$,
associated with $b_k^{\dagger}$ in Eq.~(\ref{supercharge}). 
Consequently, some of the supersymmetric transformations
that, as found above, constitute an invariant supersymmetry of the 
Hamiltonian $H'_{\rm B}$ may turn to be nonphysical. 

To examine this question, let us introduce a physical supercharge 
operator
\begin{equation}
P_{\rm ph}Q^{\dagger}P_{\rm ph}={\tilde Q^{\dagger}}P_{\rm ph}
\label{phys2}
\end{equation}
where
\begin{equation}
{\tilde Q^{\dagger}}=\sum_{k\neq l}b_k^{\dagger}{\hat\alpha}_{kl}
P_{n_k=0}P_{{\cal N}_k=0}+\sum_k b_k^{\dagger}{\hat\alpha}_{kk}P_{n_k=0}
\ .
\label{superch_phys}
\end{equation}
Similarly, we define 
\begin{equation}
{\tilde H}_{\rm B}=\sum_kE_k{\cal N}_k+\sum_k(2E_k
-G_{kk})n_k-\sum_{k\neq l}G_{kl}b_k^{\dagger}b_lP_{n_k=0}P_{{\cal N}_k
=0}\ ,
\label{Hphy}
\end{equation}
that represents another form of the Hamiltonian (\ref{Hbos}). 
The condition for the commutation of the supercharge with the 
Hamiltonian in the physical space then reads as 
$[{\tilde H}_{\rm B},{\tilde Q^{\dagger}}]P_{\rm ph}=0$,
which yields the following operator equality,
\begin{eqnarray}
& &
\sum_k(E_k-G_{kk})b^{\dagger}_k{\hat\alpha}_{kk}
+
\sum_{k\neq l}(2E_k-E_l-G_{kk})
b^{\dagger}_k{\hat\alpha}_{kl}P_{n_k=0}P_{{\cal N}_k=0}
\nonumber\\
& - &
\sum_{k\neq l}G_{kl}b^{\dagger}_k{\hat\alpha}_{ll}
P_{n_k=0}P_{{\cal N}_k=0}
-
\sum_{k\neq l}G_{kl}b^{\dagger}_k{\hat\alpha}_{lk}P_{n_l=0}
P_{{\cal N}_l=0}
\nonumber\\
& + & 
\sum_{k\neq l\neq m}G_{kl}b^{\dagger}_k{\hat\alpha}_{lm}
(P_{n_l=1}-P_{n_l=0})P_{n_k=0}P_{{\cal N}_l=0}P_{{\cal N}_k=0}
\nonumber\\
& - &
\sum_{k\neq l\neq m}G_{kl}b^{\dagger}_kb^{\dagger}_mb_l
{\hat\alpha}_{mk}P_{n_m=0}P_{{\cal N}_m=0}
\ =\ 0\ ,
\label{condit}
\end{eqnarray}
to be valid in the physical space.
It is not difficult to see that Eq.~(\ref{condit}) can only be satisfied
by imposing the above SUSY conditions for supercharges with vanishing 
nondiagonal $q_{kl}$ and ${\bar q}_{kl}$ elements, see Eq.~(\ref{susy}) 
and the text above.
The Hamiltonian (\ref{susy}) is the {\em only\/} invariant-SUSY
Hamiltonian with supermultiplet states contained entirely in the
physical space that can be constructed through a bilinear supercharge
in Eq.~(\ref{supercharge}).
The other solutions, see Eq.~(\ref{det}), are spurious, i.e., 
have no real counterpart on the level of the original fermion
system.

Let us finally note that the search for invariant supersymmetries
would be much more involved for the mapping performed via collective
bifermions from Eq.~(\ref{bifer2}). 
In this case, we again have to check if the expression 
$[{\tilde H}_{\rm B},{\tilde Q^{\dagger}}]P'_{\rm ph}$ vanishes, where 
$Q^{\dagger}=\sum_{\mu,k}c^{\dagger}_{\mu}(q_{\mu k}\alpha_k+
{\bar q}_{\mu k}\alpha_{\bar k})$ is a collective bilinear supercharge and 
$P'_{\rm ph}$ a projector onto the physical subspace obtained via 
the collective superalgebra.
Unlike the noncollective case, boson-boson and boson-fermion 
interactions in the Hamiltonian ${\tilde H}_{\rm B}$ are generally
relevant also in the physical subspace \cite{Cejnar1} and must be 
considered in the commutator.
Moreover, the calculation is obscured by the fact that $P'_{\rm ph}$ 
has a more complicated form in the general case than for the 
noncollective superalgebra.
The above-discussed relation between collective and noncollective
bosons, see the end of Sec.~\ref{mapp2}, indicates that the use of 
collective supercharges has a similar effect as the generalization of
Eq.~(\ref{supercharge}) to the form (\ref{gensup}).

\section{Conclusions}
\label{conclusions}

We have pointed out the need to distinguish between the
well-established notion of dynamical supersymmetry in many-body
physics (nuclear structure in particular), and invariant supersymmetry
in many-body physics as a direct analogue of SUSY quantum mechanics.

In contrast to phenomenological studies based on the IBFM, our
search for the SUSY patterns is derived from purely fermionic level.
Existing boson-fermion mapping techniques allow us to introduce the
relevant boson and fermion degrees consistently, and facilitates our
exploration of invariant supersymmetry in the simple case of
pairing-like Hamiltonians.

For these Hamiltonians, and with supercharges restricted to a bilinear
form, it turns out that invariant supersymmetry is ruled out for
pairing other than the simplest diagonal form, see Eq.~(\ref{susy}). 
This is not in contradiction with the result of Ref.\cite{Jolos} 
where the boson interaction does not follow from a mapped fermion 
interaction.

From a microscopic point of view, considering and accounting for the
physical subspace, has been crucial for a complete analysis, as shown
in detail in subsections \ref{mapp1} and \ref{invariant}.

Although we have not exhausted all possibilities that may result in
invariant supersymmetry for generalized pairing in a fermion system,
going beyond the present analysis seems to present a tough problem
when pursued analytically.

Finally, as argued in greater detail in Ref.\cite{CGS11}, and as is
also clear from the present discussion, there is no reason to expect
(as, e.g., in Ref.\cite{Barea}) that transfer operators in the context
of dynamical supersymmetry, should be restricted to the simplest
supercharges appearing in invariant supersymmetry situations.  In
fact, for dynamical symmetry the appropriate transfer operators are
{\it mapped images} of single fermion operators, as, e.g., in
Eqs.~(\ref{map4}) and (\ref{map5}).  In our presentation these images
may indeed be considered generalized supercharges, but in general they
are not of the simple bilinear type.

\acknowledgments{P.C. thanks J.~Dobe\v s, J.~Ho\v sek, and
D.~Nosek for interesting discussions.
This work was supported by the S.A. National Research Foundation
under Grant Nos. GUN 2047181 and GUN 2044653 and by the Grant
Agency of Czech Republic under Grant No. 202/02/0939.}

\end{document}